\pdfoutput=1

\documentclass[11pt]{article}

\usepackage[final]{acl}

\usepackage{times}
\usepackage{latexsym}

\usepackage[T1]{fontenc}

\usepackage[utf8]{inputenc}

\usepackage{microtype}

\usepackage{inconsolata}

\usepackage{graphicx}

\title{LoRP-TTS: Low-Rank Personalized Text-To-Speech}

\author{\L{}ukasz Bondaruk, Jakub Kubiak\\
  Samsung R\&D Institute Poland\\
  Plac Europejski 1\\ 00-844 Warszawa, Poland\\
  \texttt{\{l.bondaruk, j.kubiak3\}@samsung.com} \\}

\begin{document}
\maketitle
\begin{abstract}
Speech synthesis models convert written text into natural-sounding audio. While earlier models were limited to a single speaker, 
recent advancements have led to the development of zero-shot systems that generate realistic speech from a wide range of speakers 
using their voices as additional prompts. However, they still struggle with imitating non-studio-quality samples that differ significantly 
from the training datasets. In this work, we demonstrate that utilizing Low-Rank Adaptation (LoRA) allows us to successfully use even single 
recordings of spontaneous speech in noisy environments as prompts. This approach enhances speaker similarity by up to $30pp$ while 
preserving content and naturalness. It represents a significant step toward creating truly diverse speech corpora, that is crucial in all speech-related tasks.
\end{abstract}

\section{Introduction}

Recent advancements in natural language processing, particularly through large language models (LLMs), 
have significantly improved the diversity and quality of generated textual data \cite{llm}.
These innovations have led to the development of powerful systems capable of producing text across a wide range of linguistic styles and domains \cite{llm2}.
However, an equivalent focus has yet to be fully directed toward addressing speech variety.

Meeting the need for diverse and representative speech corpora remains a significant challenge,
especially as the field expands to support a broader range of languages and dialects.
High-quality, multi-speaker datasets are crucial for systems like text-to-speech (TTS), automatic speech recognition (ASR),
and speaker identification to provide a consistent user experience for people globally \cite{diversity}.
However, many datasets lack sufficient variety in linguistic features and speaker representation,
particularly in under-resourced languages, where the number of available speakers is limited.

This shortage poses a substantial obstacle to the development of robust speech-based systems.
Although significant advancements have been made in zero-shot voice cloning models \cite{valle, xtts}, which are key for enhancing diversity, 
these systems still face challenges in generalization. Despite being trained on large datasets, they often struggle with speakers whose vocal characteristics, 
speech prosody, or environment differ significantly from the training data. Furthermore, many evaluations focus primarily on 
studio-quality recordings \cite{vctk}, which may not fully capture real-world variability, highlighting the limitations of current approaches.

\section{Preliminary studies}




\textbf{LoRA} \cite{lora} is efficient finetuning method. Instead of updating all model weights, LoRA inserts low-rank matrices into specific layers of the model.
These matrices capture task-specific adjustments while keeping the majority of the original model parameters fixed.
This reduces the computational cost and memory requirements for fine-tuning, making it especially useful for large TTS models. 
It introduces two key hyperparameters: $r$ and $\alpha$. The rank $r$ controls the dimensionality of the low-rank matrices,
determining how much capacity the adaptation has to model the new task or voice.
The scaling factor $\alpha$ balances the influence of the low-rank matrices by controlling how much the model
output is impacted by the task-specific adjustments. \\
\textbf{Voicebox} \cite{voicebox} is a leading state-of-the-art zero-shot voice cloning pipeline. It is consisted of several components like: force aligner, DurationPredictor
\cite{voicebox}, Voicebox itself, and vocoder. It has superior performance in generating high-quality and coherent voice clones across various tasks.

\section{Method}

Our approach called LoRP employed finetuning voice cloning TTS with LoRA during inference time. Before using audio sample as a 
prompt to TTS, additional $100$ optimizer steps of finetuning are performed. 
For training synthesis pipeline, we utilized two types of datasets: a multilingual dataset for unsupervised pretraining and a single-language Polish dataset
for fine-tuning, ensuring robust model adaptation to diverse linguistic scenarios. We've set LoRA hyperparameters $r$ and $\alpha$ to $16$ and $16$ respectively, 
and inserted LoRA after every dense layer of Voicebox, resulting in $10M$ additional parameters, which is $2.3\%$ of all Voicebox weights.
This approach enhanced the speech similarity of synthesized voices, maintained content, allowed for the utilization 
of a minimal amount of data, added low computational overhead, and demonstrated strong generalization capabilities.

\subsection{Metrics}

\textbf{Correctness and intelligibility} refers to how accurately the synthesized speech aligns with the intended content,
ensuring that the spoken words match the original text. It also reflects the clarity and coherence of the audio output.
It might be measured by metrics such as Word Error Rate (WER) and Character Error Rate (CER).
We employed the Whisper-large-v3 \cite{whisper} model to calculate WER and CER,
which helped us evaluate both the accuracy of the synthesized speech and its intelligibility. \\
\textbf{Coherence} in the context of text-to-speech (TTS) synthesis,
refers to the similarity between the prompt audio and the synthesized output,
ensuring that the generated speech maintains consistency with the characteristics of the original speaker.
To measure coherence in our work, we utilized TitaNet-Large \cite{titanet},
to extract embeddings from both the prompt and synthesized audio samples.
These embeddings capture speaker-specific features, allowing us to compute the cosine similarity between the two.
Higher cosine similarity values indicate a closer match in the speaker's voice and style between the prompt and the synthesized speech. \\
\textbf{Quality} was measured by the commonly used Mean Opinion Score (MOS),
which provides a subjective measure of the perceived quality of audio.
We calculated an automatic MOS using the SpeechMOS \cite{speechmos} model for all our experiments,
which provided a quick and scalable assessment of the audio quality.


\subsection{Evaluation data} \label{evaluation-data}

For the evaluation, we opted for a diverse set of corpora to thoroughly test the method's potential.
In the initial experiments, aimed at gaining insights and demonstrating the approach's viability,
we used a self-compiled dataset featuring the voice of Kretes,
a character from the Reksio children's series~\footnote{\url{https://pl.wikipedia.org/wiki/Reksio_(serial_animowany)}}.
Although seemingly unconventional, this dataset contains approximately $2$ hours of speech from a single speaker,
captured in various environments and intonations, with distinct characteristics that significantly differ from the training data.
It is notably exaggerated, resembling the expressive and theatrical tones often associated with characters in role-playing games.

For the final evaluation, where we assessed the method's generalization capabilities,
we expanded our testing to include the test subset of Clarin \cite{clarin},
known for its high-quality, studio recordings, the test subset of Fleurs \cite{fleurs},
a more diverse dataset commonly used in ASR and translation tasks to represent real-world conditions,
and Nemo \cite{nemo}, a dataset focused on emotional expression across a small group of speakers.
This diverse set of corpora provided a comprehensive evaluation of our approach's adaptability and robustness.

In all experiments, the texts used for synthesis were sourced from the CommonVoice16 dataset \cite{commonvoice}.
This choice helped eliminate bias toward any specific speech corpus while providing a wide range of diverse, general-purpose sentences.

\section{Results}



We aimed to determine the minimum number of data samples required
per speaker to effectively adapt LoRA for speaker-specific voice synthesis.
We evaluated this across two dimensions: the number of samples and the number of fine-tuning steps.

First, we selected $1$, $2$, $5$, and $10$ samples from the Kretes dataset
with each sample being approximately $3$ seconds in duration,
and fine-tuned the LoRA model for $10$, $25$, $50$, $100$, and $1000$ steps.
For each fine-tuned LoRA, we synthesized $1000$ texts from the CommonVoice16
dataset using the same sample that was initially selected in the $1$ sample setup as the prompt for the synthesis. 
This allowed us to observe how varying the amount of training data and
fine-tuning steps influenced the model's ability to adapt to the speaker's voice.
To provide a baseline for comparison, we also tested the performance of Voicebox without any additional fine-tuning,
representing a general-purpose synthesis model.

Lastly, we evaluated the upper limit of LoRA's performance by
fine-tuning it on the entire speaker dataset for $3200$ steps.
This gave us insight into how much fine-tuning was necessary
to achieve optimal adaptation and how LoRA's performance scaled with increased data and training time.
These experiments helped identify the trade-offs between data availability and fine-tuning efforts for speaker-specific TTS adaptation.

\begin{figure}[h!]
    \centering
    \includegraphics[width=0.4\textwidth]{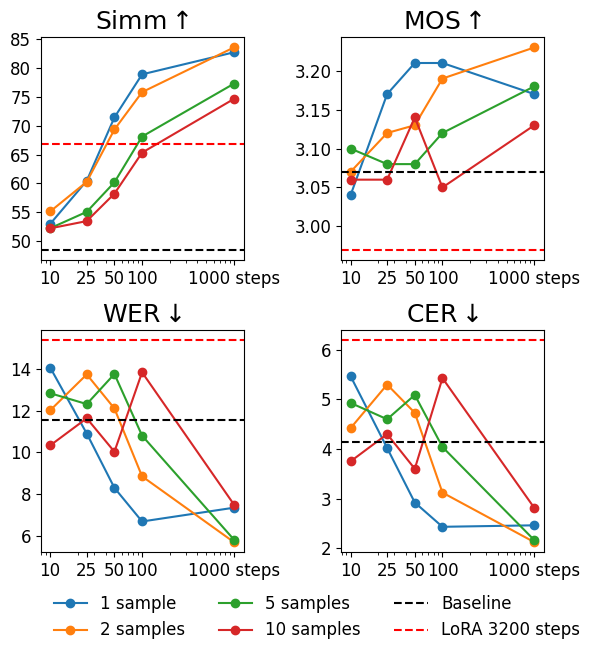}
    \caption{Evaluation metrics for various numbers of samples and optimizer steps on Kretes dataset.}
    \label{fig:eval-samples}
\end{figure}

The results presented in Figure~\ref{fig:eval-samples} reveal several compelling insights.
Notably, even a few adaptation steps yield a significant increase in speaker similarity.
As the number of steps grows, this improvement continues, with substantial gains observed with just $100$ optimizer steps.
Initially, there is a slight decline in intelligibility, as indicated by the WER and CER metrics,
likely due to the randomized state of the LoRA weights.
However, as training progresses, these metrics improve,
suggesting that Voicebox becomes more adept at capturing speaker characteristics and generating clearer,
more natural synthesis.

Furthermore, adding more data samples from the same speaker does not necessarily guarantee better performance.
Particularly when the data comes from challenging environments,
is automatically gathered, and potentially contains errors,
the overall synthesis quality may suffer. In fact, when we utilized all samples from the Kretes dataset,
we observed poorer results compared to using just one sample as seen in Figure \ref{fig:eval-samples}.
One can argue it is because of the inconsistency
in quality between samples but we find it quite usual in real-world scenarios.

As a result, we selected a setup utilizing only a single sample, optimized for $100$ steps,
which we will refer to as LoRP (Low-Rank Personalization) throughout the remainder of the study.

\subsection{Generalization}


In this follow-up experiment, we investigated the generalization capabilities of our LoRA-based method across different datasets.
We randomly sampled $100$ audio samples from each of the four datasets described in~\ref{evaluation-data}.
Using the fine-tuning parameters established in previous experiments,
we fine-tuned LoRA for each individual sample for $100$ steps, a method we refer to as LoRP.
After fine-tuning, we synthesized $100$ texts from CommonVoice16 for each LoRA model, using the corresponding fine-tuned sample as the prompt.

For comparison, we generated the same set of $100$ texts using the baseline model without any fine-tuning for each sample.
This allowed us to assess the difference between LoRP and the baseline performance.

Additionally, we conducted a third experiment where LoRA was fine-tuned on all $100$ samples from each dataset for $3200$ steps.
After fine-tuning, we synthesized the same $100$ texts for each prompt,
enabling us to evaluate the model's performance when fine-tuned on a larger set of data.
This comprehensive comparison across LoRP, baseline, and multi-sample fine-tuning helped us assess the method's generalization capabilities.

\begin{figure}[h!]
    \centering
    \includegraphics[width=0.4\textwidth]{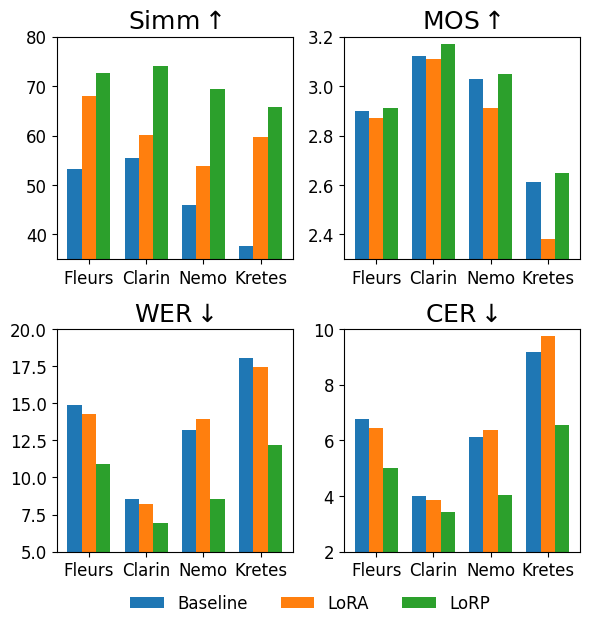}
    \caption{Evaluation metrics across different datasets.}
    \label{fig:eval-others}
\end{figure}

As demonstrated in Figure~\ref{fig:eval-others}, the insights from our earlier experiment hold.
Both Clarin and Fleurs datasets, which already exhibit good baseline speaker similarity,
showed further improvements with the application of classic LoRA and our LoRP approach.
Notably, while LoRA achieved satisfactory gains in Fleurs, its impact on Clarin was marginal.
In contrast, LoRP delivered exceptional results across all datasets.

For datasets with lower baseline similarity, such as Nemo and Kretes, both methods yielded lower absolute scores.
However, when viewed relatively, the improvements were even more impressive.
Starting from a baseline with only a slight resemblance to the voice prompt,
LoRP was able to elevate the output to high-quality synthesis.
Although LoRA also made progress, its similarity scores were lower,
and its MOS ratings fell short, suggesting that while the samples may have sounded similar, they lacked naturalness.

Finally, examining intelligibility metrics, we observed that LoRA produced some variability in WER and CER,
and produced non-reliable results, sometimes improving intelligibility and sometimes decreasing. However,
LoRP consistently improved both metrics, achieving notable gains.

\section{Ablation Study}

In this section, we analyze the impact of different configurations and methods on the performance of our system.
Two aspects were explored: the trade-off between increasing inference steps versus using our LoRP method for speaker adaptation,
and the impact of varying LoRA configurations, specifically the hyperparameters $r$ and $\alpha$, on the overall system performance.

\subsection{Inference Steps vs. LoRP Method}

Improving one dimension of a system often results in trade-offs, particularly between time and quality.
In voice cloning, the LoRP method we introduced increases the time required for inference,
but yields far greater improvements than simply increasing the number of inference steps in the Voicebox model.

We began by investigating whether increasing the number of inference steps in Voicebox,
which solves a differential equation \cite{voicebox}, could lead to better quality synthesis. As shown in Table \ref{eval-inference-steps},
this approach provided minor improvements in WER and CER, but caused a significant drop in speaker similarity,
an essential metric for voice cloning. Additionally, the MOS showed little to no improvement,
despite the increased computational time.

In contrast, the LoRP method, which fine-tunes LoRA for each speaker using a single sample,
yielded much better results in speaker adaptation and generalization. Although LoRP also increases the inference time,
its ability to personalize the synthesis for a new speaker proved far superior to merely increasing inference steps.
Thus, the LoRP method offers a more effective solution, achieving higher quality, intelligibility, and coherence in voice synthesis.

\begin{table}[h!]
  \centering
  \begin{tabular}{rrrrr}
    \hline
    \textbf{Steps} & \textbf{Simm} $\uparrow$ & \textbf{WER} $\downarrow$ & \textbf{CER} $\downarrow$ & \textbf{MOS} $\uparrow$ \\
    \hline
    $15$ & \textbf{48.00} & $14.06$ & $5.47$ & $3.04$ \\
    $30$ & $43.52$ & \textbf{11.75} & \textbf{4.17} & \textbf{3.05} \\
    $45$ & $43.54$ & $11.80$ & $4.24$ & \textbf{3.05} \\
    $60$ & $43.52$ & $12.00$ & $4.35$ & \textbf{3.05} \\
    \hline
  \end{tabular}
  \caption{\label{eval-inference-steps} Evaluation metrics for various inference steps.}
\end{table}

\subsection{LoRA Configuration}

In this part of the study, we examined how varying the LoRA hyperparameters, $r$ and $\alpha$, affected the model's
performance in terms of speaker similarity, intelligibility, and overall quality.
We synthesized $10k$ sentences for each configuration, fine-tuning the Voicebox model using
LoRA for $3200$ optimizer steps on the full Kretes dataset.

\begin{table}[ht]
\centering
\begin{tabular}{rrrrrr}
\hline
\textbf{$r=\alpha$} & \textbf{Simm} $\uparrow$ & \textbf{WER} $\downarrow$ & \textbf{CER} $\downarrow$ & \textbf{MOS}  $\uparrow$ \\ \hline
$4$ & $59.79$ & \textbf{16.36} & \textbf{8.32} & \textbf{2.44} \\
$8$ & $59.94$ & $18.68$ & $9.81$ & $2.28$ \\
$16$ & $59.78$ & $17.44$ & $9.76$ & $2.38$ \\
$32$ & \textbf{61.34} & $17.48$ & $9.15$ & $2.43$ \\
$64$ & $60.05$ & $18.04$ & $9.51$ & $2.24$ \\ \hline
\end{tabular}
\caption{\label{lora-params}Evaluation of various configurations of LoRA $r$ and $\alpha$ hyperparameters}
\end{table}

\begin{table*}[ht!]
  \centering
  \begin{tabular}{lrrrrrrr}
    \hline
    \textbf{DATASET\textbackslash LANGUAGE} & \textbf{pl} & \textbf{en} & \textbf{de} & \textbf{es} & \textbf{fr} & \textbf{it} & \textbf{others} \\
    \hline
    Internal & $904$ & $592$  & $2974$ & $2516$ & $2727$ & $2815$ & $1807$ \\
    CommonVoice16 \cite{commonvoice} & $173$ & $3394$ & $1394$ & $2188$ & $1105$ & $386$ & $892$ \\
    LibriLight \cite{librilight} & $0$   & $16665$ & $0$ & $0$ & $0$ & $0$ & $0$ \\
    MLS \cite{mls} & $107$ & $41678$& $1893$ & $907$  & $1054$ & $254$  & $1545$ \\
    VCTK \cite{vctk} & $0$   & $82$   & $0$    & $0$    & $0$    & $0$    & $0$ \\
    Voxceleb1 \cite{voxceleb} & $0$   & $306$  & $0$    & $0$    & $0$    & $0$    & $0$ \\
    Voxceleb2 \cite{voxceleb2} & $0$   & $2153$ & $0$    & $0$    & $0$    & $0$    & $0$ \\
    Voxlingua \cite{voxlingua107} & $80$  & $48$   & $39$   & $38$   & $66$   & $50$   & $266$ \\
    Voxpopuli \cite{voxpopuli} & $4945$& $5943$ & $5767$ & $5461$ & $5362$ & $5228$ & $12512$ \\
    \hline
    \textbf{Total} & \textbf{6209} & \textbf{70861} & \textbf{12067} & \textbf{11110} & \textbf{10314} & \textbf{8733} & \textbf{17022} \\
    \hline
  \end{tabular}
  \caption{\label{datasets-pretraining} Pretraining dataset sizes in hours per language. Other languages include ru, ro, ja, sv, and nl}
\end{table*}

As shown in Table \ref{lora-params}, the differences in performance across various configurations were minimal.
We found that the configuration $r = 16$ and $\alpha = 16$ offered the best balance between speaker similarity,
intelligibility, and resource efficiency

In conclusion, our ablation study demonstrates that while increasing inference steps offers marginal improvements,
the LoRP method and optimal LoRA configurations provide a more robust and efficient approach for voice cloning and adaptation tasks.

\section{Related works}

Previous studies \cite{tlsvmttss, speechadapt1, speechadapt2} have demonstrated methods for quickly adapting 
TTS models use only a few examples. However, these approaches are mostly evaluated on high-quality, 
studio-like corpora, which may not effectively represent real-world conditions. They also exhibit a significant 
drop in performance when the voice sample is limited to only a few seconds.

In similar situations in computer vision, LoRA has led to notable improvements. It is a fine-tuning method 
that adds trainable low-rank matrices to the original model weights. It has enabled the adaptation of text-to-image 
diffusion models to new styles with as few as five images \cite{cvlora, cvlora2}.

Initial attempts to apply LoRA to voice cloning \cite{audiobox} produced suboptimal results due to the ambitious goal 
of learning an entirely new language, but they laid the groundwork for future progress. Later research \cite{voiceboxadapt} 
demonstrated LoRA's effectiveness in adapting models to more specific domains, such as laughter and punctuation.

\section{Experimental details}


\subsection{Pretraining data}

During the pretraining phase, we gathered approximately $137k$ hours of audio recordings across $11$ languages.
As shown in Table~\ref{datasets-pretraining} in addition to open-source datasets,
we used our internal datasets.

For datasets like LibriLight \cite{librilight} and VoxPopuli \cite{voxpopuli},
which contain long recordings, we segmented the audio based on silence into shorter fragments ranging from $5$ to $20$ seconds.
To prevent overpopulating the corpus with data from a single source, we used only $3$ out of $12$ available years of VoxPopuli \cite{voxpopuli}
recordings: $2018$, $2019$, $2020$. To optimize training efficiency, we excluded any remaining records longer than $20$ seconds.
Audio signals were normalized to a $16$ kHz sample rate and converted to single-channel.
The waveforms were then transformed into $80$-dimensional log-scaled Mel spectrograms using a hop size of $256$,
window length of $1024$ and frequency from $0k$Hz to $8k$Hz.

\begin{table*}[ht!]
  \centering
  \begin{tabular}{lrr}
    \hline
    \textbf{Dataset} & \textbf{Samples} & \textbf{Duration [h]} \\
    \hline
    Internal & $447774$ & $561.62$ \\
    CommonVoice16 \cite{commonvoice} & $84375$ & $104.78$ \\
    Wikimedia \cite{wikimedia} & $17170$ & $34.28$ \\
    Mailabs \cite{mailabs} & $11164$ & $20.75$ \\
    MCspeech \cite{mcspeech} & $20512$ & $18.42$  \\
    MLS \cite{mls} & $4531$ & $17.87$  \\
    Fleurs \cite{fleurs} & $1716$ & $5.15$ \\
    \hline
    \textbf{Total} & \textbf{587242} & \textbf{762.87} \\
    \hline
  \end{tabular}
  \caption{\label{dataset-finetuning} Fine-tuning dataset samples and duration in hours.}
\end{table*}

\subsection{Fine-tuning data}

For fine-tuning, we concentrated on the Polish language
by gathering a corpus of audio recordings paired with corresponding transcriptions.
To ensure transcription quality, we applied a filtering process based on ASR verification.
We employed the Whisper \cite{whisper} model to generate transcriptions and discarded any samples where
the ASR transcription differed from the actual transcription.
The preparation of audio recordings followed the same procedure as in the pretraining phase,
with normalization applied to all texts. Additionally,
we phonetized the normalized texts using an internal Grapheme-to-Phoneme (G2P) tool,
ensuring consistency in the phonetic representation.

\subsection{Synthesis pipeline}


Voicebox \cite{voicebox} is a model designed to perform zero-shot audio reconstruction
from textual inputs and speech prompts. Voicebox operates on Conditional Flow Matching (CFM) \cite{flowmatching},
a process that transitions from Gaussian noise to a targeted audio distribution by identifying the vector's
directional flow within the audio.

The model's architecture is an adaptation of the transformer \cite{transformer} architecture,
incorporating rotary positional embedding instead of convolutional positional embedding as in the original Voicebox.
Another architectural change is the lack of ALiBi self-attention bias \cite{alibi}.
An open-source implementation of Voicebox\footnote{\url{https://github.com/lucidrains/voicebox-pytorch}}
was used to facilitate this research.

For aligning input tokens with audio features,
we utilized the DurationPredictor \cite{voicebox} model,
a smaller variant of Voicebox. This model predicts token durations based on contextual information,
such as prompt tokens and their durations. Unsupervised token duration discovery was achieved through CTC-based forced alignment.
A transformer model was trained using connectionist temporal classification (CTC)
loss to predict token labels for each audio frame,
after which a forced alignment algorithm was applied to extract token durations from aligned sequences.

To map audio features to speech, we employed the HiFi-GAN \cite{hifigan} vocoder,
which consists of a fully convolutional generator and two discriminators—multi-scale and multi-period.
HiFi-GAN was trained adversarially to reconstruct raw waveforms from the provided audio features,
producing high-quality audio outputs.

\begin{figure}[h!]
  \centering
  \includegraphics[width=0.52\textwidth]{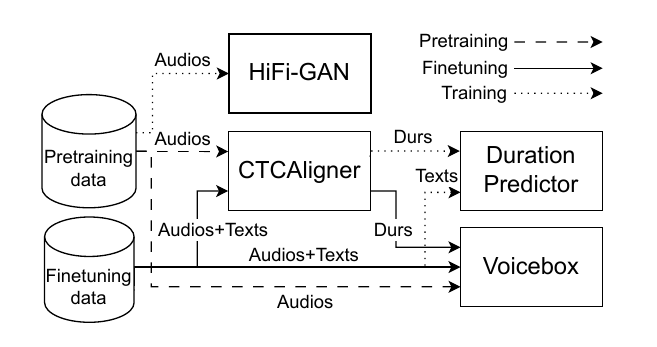}
  \caption{Synthesis training pipeline.}
  \label{fig:voicebox-training-pipeline}
\end{figure}

The entire training pipeline is depicted in Figure~\ref{fig:voicebox-training-pipeline}.

\section{Conclusions}

In this work, we proposed utilizing LoRP in voice cloning systems as a solution for synthesizing low-resource speakers. 
Our approach showed notable improvements in speaker similarity and content naturalness, even when adapting to voices with 
minimal data, such as a single sentence. Across multiple datasets, LoRP consistently excelled, particularly in challenging 
cases where baseline similarity was low. Given its efficiency, LoRP presents significant commercial potential by offering 
a cost-effective solution for reducing manual recording efforts and mitigating issues with low-quality synthesis. Looking ahead, 
we aim to expand the application of LoRP to cross-lingual setups, allowing richly diverse languages to enhance variability in 
another language, further broadening its versatility and driving continued advancements in low-resource speech synthesis.
Another area that we want to work with is intonation and emotional expression.
We want to achieve ability to capture and transfer nuances like whispering,
questioning, or laughing across different voices. It would be an exciting advancement.
However, we acknowledge that this presents a challenging task.
\bibliography{acl_latex}

\end{document}